\documentclass[5p]{elsarticle} % uncomment if you want to see the final style

\usepackage{lineno}
\modulolinenumbers[5]

\journal{NIM A}
\usepackage{graphicx,xcolor,textpos,tabularx}
\usepackage{epstopdf}
\usepackage{amsmath}
\usepackage{xcolor}
\usepackage{import}
\usepackage{graphicx} % Include figure files
\usepackage{dcolumn}  % Align table columns on decimal point
\usepackage{bm}       % bold math
\usepackage{amssymb}
\usepackage[colorlinks=true, allcolors=blue]{hyperref}
\usepackage{cuted}%needed to extend a forumla over two columns

\biboptions{longnamesfirst,sort&compress}
\begin{document}
\graphicspath{{./Figures/}}
\begin{frontmatter}

\title{Enhancing the mass resolving power of FRIB's proposed high-voltage MR-ToF mass separator and spectrometer: addressing non-ideal conditions}

% AUTHORS

\author[FRIB,MSU]{C.~M.~Ireland}
\corref{mycorrespondingauthor}
\cortext[mycorrespondingauthor]{Corresponding author}
\ead{ireland@frib.msu.edu}
\author[FRIB]{F.~M.~Maier} 
\author[FRIB,MSU]{E.~Dhayal}
\author[BerkeleyAddress]{E.~Leistenschneider}
\author[FRIB,MSU]{R.~Ringle}
\author[FRIB,MSU2]{A.~Sjaarda}

% AFFILIATIONS
\address[FRIB]{Facility for Rare Isotope Beams, East Lansing, Michigan 48824, USA}
\address[MSU]{Department of Physics and Astronomy, Michigan State University, East Lansing, Michigan 48824, USA}
\address[BerkeleyAddress]{Nuclear Science Division, Lawrence Berkeley National Laboratory, Berkeley, California 94720, USA.}
\address[MSU2]{College of Engineering, Michigan State University, East Lansing, Michigan 48824, USA}

\begin{abstract}
 Multi-reflection time-of-flight mass separators and spectrometers (MR-ToF MSs) are indispensable tools at radioactive ion beam (RIB) facilities. These electrostatic ion beam traps act as highly selective mass separators and high-precision mass spectrometers for rare and exotic nuclei. When well-tuned and designed to minimize higher-order flight-time aberrations, state-of-the-art MR-ToF MSs approach, and slightly exceed, mass resolving powers of $m/\Delta m = 10^6$. Achieving $m/\Delta m >3 \cdot 10^6$ would provide the ability to resolve $>90$\% of all known isomeric states with half-lives above 10~ms. However, the ability to mass separate in all practical setups is limited by non-ideal conditions which place such resolving powers out of reach. To this end, we present a simulated analysis of these conditions in the newly proposed high-voltage MR-ToF MS for the Facility for Rare Isotope Beams (FRIB). It is expected to store ions at 30 keV beam energy and increase ion throughput by two orders of magnitude compared to current devices. Existing efforts to mitigate the effects of non-ideal conditions employed for current MR-ToF devices storing ions at $<3$~keV beam energy will already enable mass resolving powers approaching $10^6$ for FRIB’s high-voltage MR-ToF device. Simulations of newly proposed mitigation strategies show that even mass resolving powers approaching $10^7$ might become feasible.
\end{abstract}

\end{frontmatter}
%\linenumbers
\section{\label{sec:Introduction}Introduction}
Many experiments at radioactive ion beam (RIB) facilities, such as the Facility for Rare Isotope Beams (FRIB), rely on pure beams to perform measurements at stopped (up to 60~keV\footnote{Note that all energies stated throughout this manuscript are given for singly charged ions.}) and/or reaccelerated (300 keV/u to 12 MeV/u) energies. Delivering isobarically (same mass number) and/or isomerically (same isotope, different energy state) pure beams hinges on the ability to separate the ion species of interest from unwanted contaminants. A unique identifier for an ion species is found in its mass $m$ and the ability to distinguish between two ion species with mass $m$ and mass $m+\Delta m$ is defined by the mass resolving power $R=m/\Delta m$. The resolving powers required span a broad spectrum ($0\lesssim R\lesssim 10^7$). In response to the need for pure beams at high ion intensity, we have completed the conceptual design of a high-voltage multi-reflection time-of-flight mass separator and spectrometer (MR-ToF MS) for FRIB~\cite{FRIBMRToF}. Ion bunches prepared in an upstream Paul-trap cooler-buncher with a minimal temporal bunch width are injected into the MR-ToF device consisting of two electrostatic mirrors connected by a central drift tube. Once stored by a switching of potentials, ions undergo revolutions between the two mirrors and separate in time-of-flight by their mass-over-charge ratio $m/q$. The FRIB MR-ToF device is designed to store ions at 30 keV beam energy, with the goal of increasing the ion throughput by two orders of magnitude~\cite{MAIER2023168545, FRIBMRToF} compared to state-of-the-art low-voltage MR-ToF MSs~\cite{PLASS2013, Schury2009, Wienholtz2013, REITER2021165823, CHAUVEAU2016211,LIU2021164679, ROSENBUSCH20231e6, virtanen2025highresolutionmultireflectiontimeofflightmass} that store ions at $< 3$~keV beam energy.  These existing low-voltage devices routinely reach mass resolving powers $R>5\cdot10^5$ within only a few (tens of) milliseconds of storage time. This allows for the resolution of nearly all (molecular) isobars and nearly half of the known isomers with half-lives $>10$~ms~\cite{Dickel2024}. For $R\geq3\cdot10^6$, $>90$\% of these known isomers would be resolved~\cite{Dickel2024}. Demonstrations slightly exceeding $R=10^6$ within storage times of 22 and 12.5 ms have been performed using the MR-ToF MS at the Fragment Separator at GSI~\cite{Mardor20211e6} and at RIKEN~\cite{ROSENBUSCH20231e6}. 

\begin{figure*}[t]
\centering
\includegraphics[width=1.0\linewidth]{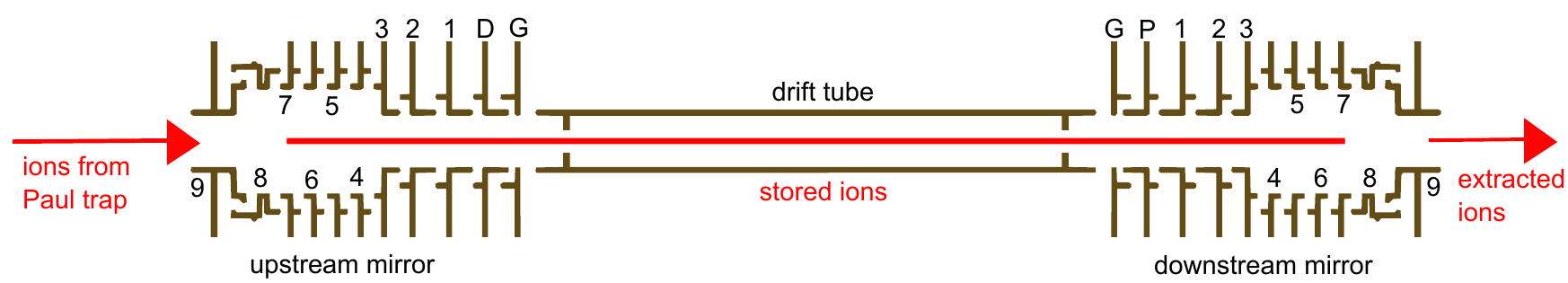}
\caption{A cross-sectional view of the simulated FRIB MR-ToF design as presented in Ref.~\cite{FRIBMRToF}. The various optical elements, including mirror electrodes 1-9, grounded electrodes (G), the deflector (D) and pickup (P) electrodes and the drift tube are labeled in black. The beam profile of the stored ions, as delivered from an upstream room-temperature buffer-gas-filled Paul-trap cooler-buncher, along with the direction of beam propagation from injection to extraction, is displayed in red.}
\label{fig:Schematic}
\end{figure*}

A goal for the next generation of these devices will be to comfortably exceed $R=10^6$ within timescales compatible with the short half-lives of the radioactive ion species of interest. Simulations show promise in reaching $R=7.5\cdot10^6$ at JINR~\cite{Yavor_2022} and $R=2\cdot10^7$ at FRIB~\cite{FRIBMRToF}, respectively. For these designs, a simulated $R=10^6$ is reached within $\approx5$~ms of storage for 1.7~ns bunch widths at JINR~\cite{Yavor_2022} and 7.4~ms for 3.7~ns bunch widths at FRIB~\cite{FRIBMRToF}\footnote{For the proposed FRIB design, shorter storage times could be possible via the dynamical time focus shift technique~\cite{DICKEL20171}. However, given the high voltages that will be present, the implementation of this technique is not foreseen during the first years of operation.}.

While these resolving powers are impressive, simulations are performed assuming ideal experimental conditions. The most crucial of these assumptions are: accurate and stable power supplies, a temperature-stable setup, perfectly aligned electrode configurations, ideal high voltage switches, and a perfect vacuum. In practice, these conditions are not met, and the mass resolving power is limited by broadening or shifts in the spectral time-of-flight distribution of stored ions~\cite{SCHURY2013537, SCHURY201439, doi:10.1063/1.5104292, WIENHOLTZ2020348, Fischer2021HVstab, LIU2021164679, Schury2009}. To assess the impact of non-ideal conditions and the effectiveness of mitigation strategies on the mass resolving power, we conducted simulations using the conceptual design for the FRIB MR-ToF MS~\cite{FRIBMRToF}. In particular, our work demonstrates that high mass resolving powers can be achieved in high-voltage, high ion-throughput MR-ToF MSs storing ions at 30~keV beam energy even when imposing non-ideal conditions. Section~\ref{sec:Design} gives a short overview of the design, detailed in Ref.~\cite{FRIBMRToF}, and introduces the simulation procedure. Section~\ref{sec:Voltage} discusses long- and short-term power supply instabilities along with existing and future mitigation techniques. Section~\ref{sec:Temperature} will do the same but for fluctuations in the ambient temperature. Section~\ref{sec:Further} will assess the impacts from voltage inaccuracies, electrode misalignments, realistic voltage switches, and the residual gas pressure. 

\section{\label{sec:Design}Design and simulation procedure of FRIB's high-voltage MR-ToF setup}
\subsection{Proposed experimental setup}
The newly proposed 30-keV MR-ToF MS setup at FRIB is discussed in detail in Ref.~\cite{FRIBMRToF}. As such, only a short summary will be provided next.

A Paul-trap cooler-buncher~\cite{PascalsPaultrap} located upstream of the MR-ToF device will prepare ion bunches with as low as $\approx4$~ns full width at half maximum (FWHM) bunch widths for injection into the MR-ToF device. When ion bunches pass through the middle of the central drift tube of the MR-ToF device, the drift tube is switched from a positive voltage to ground to efficiently trap the ions via the in-trap lift-switching technique~\cite{WOLF20128}. 
The MR-ToF device, displayed in Figure~\ref{fig:Schematic}, is $\approx1.2$~m long and consists of two axially symmetric mirrors, each containing nine electrodes, connected by a 431~mm long drift tube. Placed between the drift tube and the mirrors are grounded electrodes and deflector/pickup electrodes. While ions are stored, the azimuthally segmented deflector electrode can be utilized to remove contaminant species separated in-flight~\cite{Fischer20182} while the pickup electrode can provide feedback on the evolving bunch properties via detection of an image-charge current if ion intensities are sufficient~\cite{doi:10.1021/acs.analchem.7b02797}.  Further specifics are presented in Ref.~\cite{FRIBMRToF}, and optimal set values for the MR-ToF are shown in Table~\ref{tab:MRToFSetting} of the appendix.

The MR-ToF MS can operate in two modes, one for precision mass measurements/beam diagnostics and the other for highly selective mass separation. In the former, upon extraction of the ions via in-trap lift switching, all stored ion species are detected on a retractable detector. For the latter, the extracted ions are sent to a Bradbury Nielsen gate (BNG)~\cite{PhysRev.49.388, PLA20084560, WOLF201282} that selects the ions of interest prior to delivery to the desired experimental station within FRIB's stopped or reaccelerated beam facilities. In this work, we focus primarily on the mass separation mode.

\subsection{Simulation procedure for mass resolving power}

The mass resolving power of an MR-ToF device can be expressed according to Ref.~\cite{WOLF2013123} as

\begin{equation}\label{Eq: ResolvingPower}
R=\frac{m}{\Delta m}=\frac{t}{2\Delta t}=\frac{t_{0}+t_s+t_{d}}{2\sqrt{\Delta t^2_0+(r\Delta t_1)^2}}.
\end{equation}

Here, $t$ denotes the total flight time of the ions, while $\Delta t$ refers to the temporal bunch width (FWHM) of the extracted ions recorded at the ion detector. The total flight time is the sum of the passage time from the Paul trap to the MR-ToF device's middle plane, $t_0$, the storage time in the MR-ToF device, $t_s$, and the time taken to travel from the MR-ToF device to a downstream detector, $t_d$. The storage time is given by $t_s=rt_1$, where $r$ is the number of revolutions in the device and $t_1$ denotes the time for one revolution. The temporal bunch width, $\Delta t$, is the root sum of the squares of the initial bunch width of the ions injected into the device, $\Delta t_0$, and the temporal peak broadening, $r\Delta t_1$, where $\Delta t_1$ is this broadening per revolution. It should be noted that while the mass resolving power via Eq.~\ref{Eq: ResolvingPower} suffices for distinguishing two species of a given $m/\Delta m$ ratio, the mass resolving power needed to fully separate the ion species is some factor higher~\cite{WOLF2013123}. The mass resolving power in the limit of infinite revolutions $R_\textrm{inf}$ is given by $R_\textrm{inf}=t_1/2\Delta t_1$. 

As outlined in Ref.~\cite{FRIBMRToF}, we simulate the entire ion path from buffer-gas cooling in the Paul trap to extraction from the MR-ToF device in SimIon~\cite{SimIon}. To avoid simulating the full ion storage time, which can become computationally expensive, one can simulate a few revolutions and extract the parameters required in Eq.~\ref{Eq: ResolvingPower} from the recorded data~\cite{MAIER2023168545}. The parameters presented herein are typically recorded in the middle plane of the MR-ToF device. 

\begin{figure}[t]
\centering
\includegraphics[width=\columnwidth]{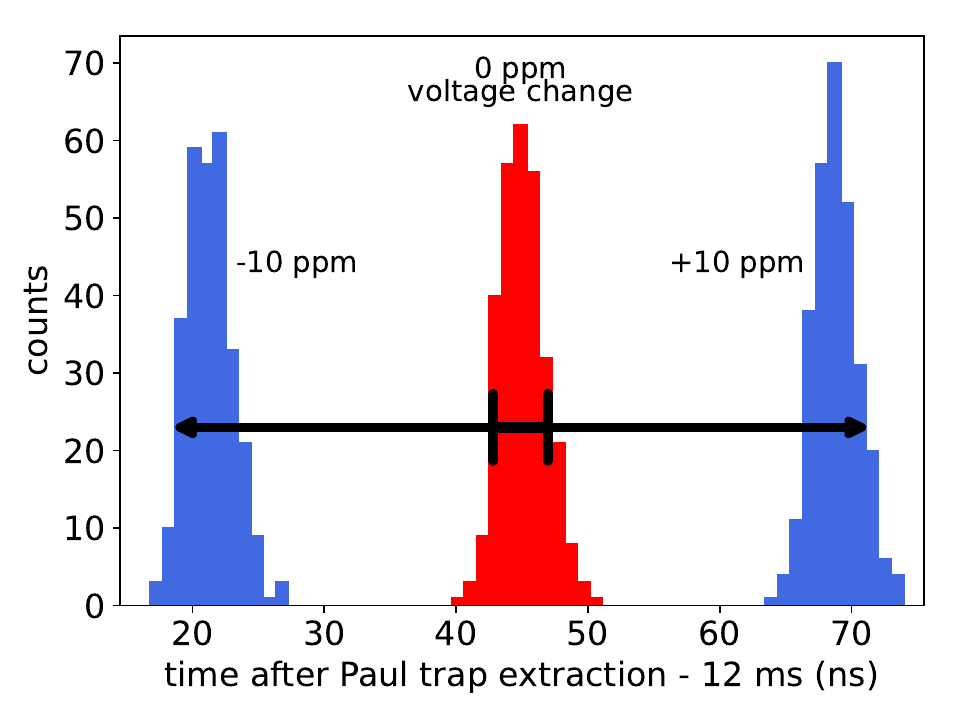}
\caption{The simulated time of flight distribution of stored ions when applying the optimal voltage (red peak) or a $\pm10$~ppm shift (blue peaks) to the outermost mirror electrode~(\#9) after $\approx12$~ms of storage time in the FRIB design. The black arrows display the window that the time centroid varies under this long-term instability, resulting in a broader spectrum on the downstream detector compared to the initial bunch width given by black brackets.}
\label{fig:Broadening}
\end{figure}

One individual measurement cycle consists of the preparation of the ion bunch in the Paul trap, storage in the MR-ToF device, and subsequent detection. In a typical MR-ToF measurement, thousands of individual measurement cycles are repeated and summed. Various instability sources discussed in Sections~\ref{sec:Voltage}--\ref{sec:Further} can result in shifts in the ion flight time, leading to an additional broadening of the ion bunch when summing the extracted ions over many individual measurement cycles that slightly differ in their conditions. For example, simulated time-of-flight spectra are provided in Figure~\ref{fig:Broadening} with -10, 0 and +10 ppm voltage shifts applied to mirror electrode \#9, in line with the expected $\pm10$~ppm long-term instability from commercially available non-stabilized high-voltage power supplies (e.g FUG HCP series). In this case, the centroid of the time-of-flight bunch varies by $\pm25$~ns, broadening the experimentally observed FWHM from $5$~ns to $50$~ns when taking multiple measurement cycles into account during which the power supply voltage on mirror electrode 9 drifted -10 ppm to +10 ppm from its desired set value. This would worsen the mass resolving power at this storage time from $R\approx 1 \cdot 10^6$ to $R\approx 1 \cdot 10^5$. 

Simulations can assess the individual contributions of various non-ideal conditions in MR-ToF devices while providing an environment to study mitigation techniques. All simulations presented herein are performed using singly charged ions of mass $m=133$~u. Some simulations are repeated for $m=24$~u and 250~u. No mass-dependent changes in performance, aside from different masses having (slightly) different initial bunch widths given by the chosen Paul trap extraction settings, are observed.

\section{\label{sec:Voltage}Voltage instabilities}
\subsection{Description and simulated impacts}
\subsubsection{\label{subsec:LT}Long-term voltage drifts}
FRIB plans to perform highly selective mass separation of rare isotopes with an MR-ToF device over long measurement times (on the order of days). Therefore, we first consider voltage fluctuations occurring on a timescale that exceeds that of a full measurement cycle. Due to such fluctuations, individual ion bunches prepared in subsequent measurement cycles are subject to slightly different mirror potentials, altering their time-of-flight centroid when extracted at a given revolution number. In this case, we are concerned not with the instantaneous bunch width $\Delta t_0$ of a single stored ion bunch, but with the drift in its flight time due to a gradual deviation from the optimal set voltage. 
As discussed above, a potential drift of $\pm 10$~ppm for our outermost mirror electrode can worsen the mass resolving power by an order of magnitude (see Fig.~\ref{fig:Broadening}). It is crucial to individually assess the sensitivity of the flight time to voltage drifts for each of the nine mirror electrodes to estimate the potential worsening in mass resolving power.

\begin{figure}[!t]
\centering
\includegraphics[width=1.0\linewidth]{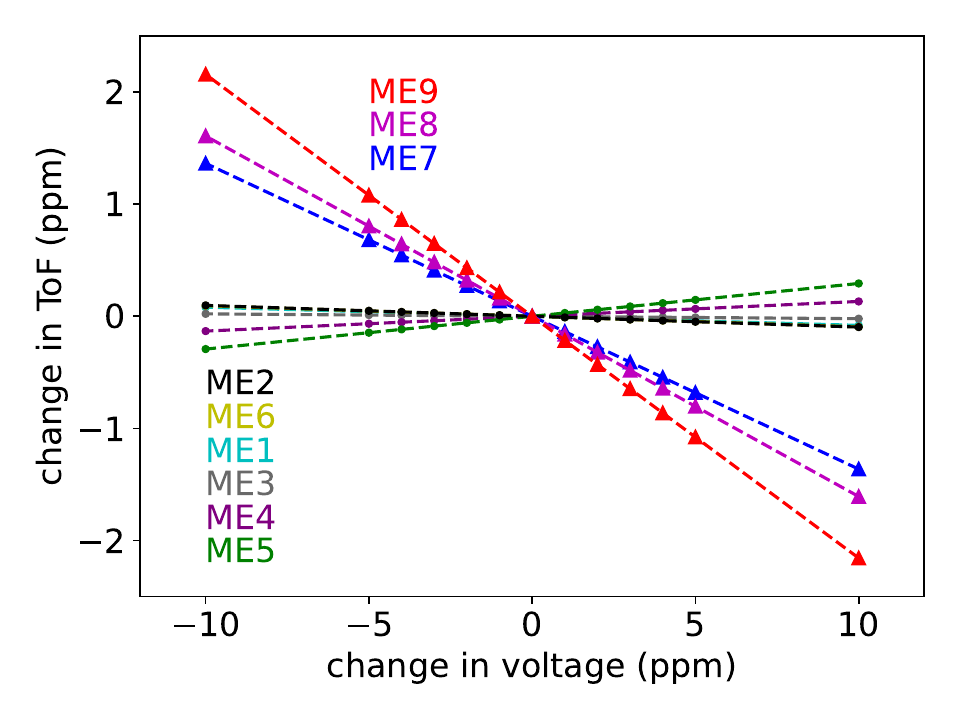}
\caption{The simulated change in time of flight (ToF) as a function of change in voltage (in units of parts-per-million ppm) for the FRIB MR-ToF device mirror electrodes (labeled vertically as they appear on the left-hand side of the plot). Each of the nine mirror electrodes (ME) individually has their set voltage varied up to $\pm10$~ppm around their optimal value. The error bars are smaller than the individual data markers.}
\label{fig:LTDrift}
\end{figure}

To simulate this, we implement the procedure initially discussed in Ref.~\cite{Schury2009}. An ion bunch consisting of 300 ions is injected into the MR-ToF device, and its time-of-flight centroid is recorded after 25 revolutions. Afterwards, we restart the simulation. With each restart, a parts-per-million shift of up to $\pm10$~ppm from the optimal set value is applied to only a single mirror electrode, such that the ions experience a different potential distribution per simulation run. This procedure allows for an approximation of the worsening in the mass resolving power using Eq.~\ref{Eq: ResolvingPower} and provides insight into the sensitivity of a given mirror electrode relative to the others. The change in the time-of-flight centroid as a function of voltage shifts for the nine mirror electrodes is displayed in Figure~\ref{fig:LTDrift}. The slope of each line corresponds to the sensitivity of the ion flight time to voltage changes for that given electrode. From this, we find that the flight time of stored ions is most sensitive to drifts in the outermost mirror electrodes \#7-9 (triangle markers), which are biased to the highest potentials (up to 43.3~kV). For example, in mirror electrode 9 (most sensitive), a 1~ppm shift in voltage corresponds to a 0.2~ppm shift in the time centroid, while in mirror electrode 3 (least sensitive), this same shift amounts to only a 0.008~ppm time-centroid shift. Therefore, mirror electrodes \#7-9 will have first priority in experimental stabilization efforts.

Since all mirror electrodes will be affected by these drifts, we must consider the total broadening to determine the expected worsening in the resolving power. We only consider a worst-case scenario; hence, no drifts in the individual voltages are canceling each other out. We find that for commercial high-voltage power supplies that are not actively stabilized ($\pm10$~ppm instability), the mass resolving power worsens from $R_\textrm{inf}=2\cdot10^7$ to $4\cdot10^4$. With the recent advancements in the active stabilization of high-voltage power supplies employing precision voltage dividers~\cite{PASSON2025101818, 10.1063/5.0218649}, $\pm0.5$~ppm stabilization has been maintained for the order of weeks for 60~kV power supplies. A $\pm0.5$~ppm voltage drift across all mirror electrodes, without any cancellations, gives $R_\textrm{inf}\approx8\cdot10^5$. This resolving power is sufficient to resolve nearly all (molecular) isobars and more than half of the known nuclear isomers with half-lives above 10~ms~\cite{Dickel2024}. For a mass resolving power of $3 \cdot 10^6$ and eventually $1\cdot10^7$, a voltage stabilization to $\pm0.125$~ppm and $\pm0.036$~ppm would be required, respectively.

Considering the FRIB MR-ToF device will be the first to employ 9 mirror electrodes, we assess the impact of the number of mirror electrodes on the stability against long-term voltage drifts. We perform similar simulation studies as shown in Fig.~\ref{fig:LTDrift} using the six mirror electrode 30~keV MIRACLS design~\cite{MAIER2023168545}, another high-voltage MR-ToF device at ISOLDE/CERN. There, a $\pm0.5$~ppm voltage drift applied to all six mirror electrodes worsens the mass resolving power from $R_\textrm{inf}\approx2\cdot10^6$ to $\approx6\cdot10^5$. We conclude that, in the study of long-term voltage drifts, no added instability is derived from our increase in the number of mirror electrodes. With current stabilization techniques for long-term voltage drifts, the FRIB design is predicted to be on par with state-of-the-art devices that approach resolving powers $R>10^6$ in a few (tens of) milliseconds~\cite{Mardor20211e6,ROSENBUSCH20231e6}. Further advancements in the stability of high-voltage power supplies and additional mitigation strategies are expected to enable even higher mass resolving powers, see Sec.~\ref{subsec:VoltageMit}. 

\subsubsection{\label{subsec:ST}Short-term voltage instabilities}
In addition to the long-term voltage drifts discussed in Section~\ref{subsec:LT}, voltage instabilities on time scales slightly larger than one revolution period, in the form of 60~Hz sinusoidal pickup, and far shorter than the revolution period, in the form of Gaussian white noise, are also present. In the following, each noise source is simulated independently using a procedure similar to the one outlined in Ref.~\cite{MAIER2023167927} to assess the impact on the mass resolving power.

\textbf{Sinusoidal noise (no phase-locking)--}
The 60~Hz sinusoidal noise due to electrical pickup originates from the ground connection of the individual mirror electrode power supplies. This amounts in simulation to adding a 60~Hz sinusoidal signal to the optimal set voltage of each mirror electrode while keeping the pickup, deflector, shielding and drift tube electrodes at ground. The amplitude of the 60 Hz sinusoidal noise is expected to be below 0.5~V. Assuming the same ground connection for all power supplies, the starting phase of this added noise and the amplitude is kept the same for all electrodes. If the ion extraction from the Paul trap is not locked to a phase of the sinusoidal noise, subsequent ion bunches will experience a different initial phase. To simulate this, we allow ions to undergo 5000 revolutions  ($\approx45$~ms of storage time for $m=133$~u) in the MR-ToF device one at a time. With each ion, the mirror electrode voltages are given a random initial phase of the sinusoidal curve between $-\pi$ and $\pi$. For storage times exceeding the period of the sinusoid (16.7~ms), the full oscillation is nicely sampled over. 

\begin{figure}[t]
\centering
\includegraphics[width=1.0\linewidth]{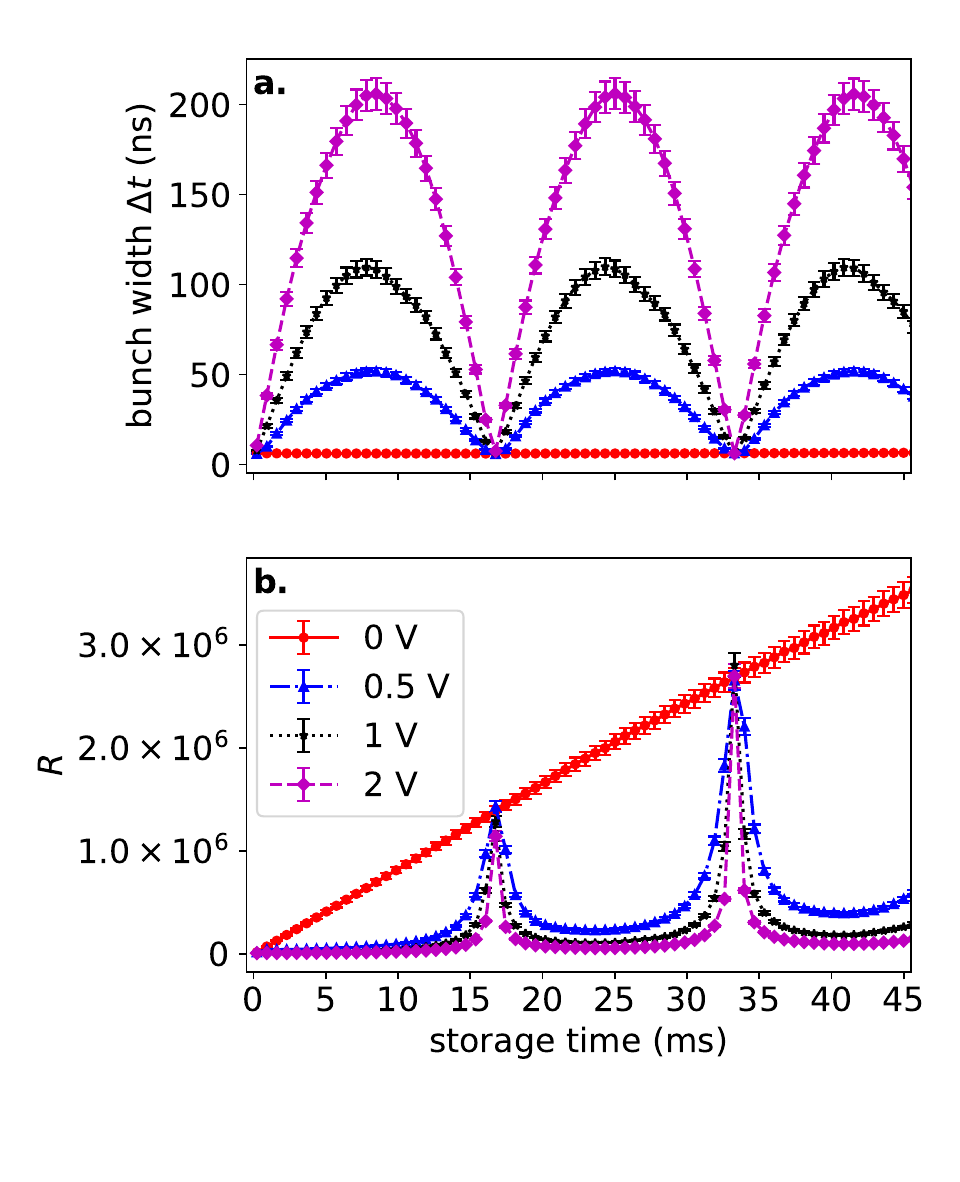}
\caption{(a.) The change in the bunch width and (b.) the mass resolving power $R$ of stored 5~ns ion bunches as a function of storage time for various 60~Hz sinusoidal noise amplitudes with a random initial phase per ion.}
\label{fig:Sinus1}
\end{figure}

Sinusoidal amplitudes of 0.5, 1.0 and 2.0~V are simulated for 5~ns initial ion bunch widths. The results of these simulations are displayed in Figure~\ref{fig:Sinus1}. As a result of the imposed sinusoidal noise with a random initial phase per ion, individual ions experience slightly different potentials when passing through the mirror electrodes and entering a grounded region of the device. The maximum difference in revolution period due to this energy change is proportional to the applied noise and becomes maximum at half the sinusoidal period. This results in a maximum broadening in the temporal bunch width and a relatively poor mass resolving power. With the completion of a period of the sinusoid ($\approx16.7$~ms for 60~Hz noise), the accumulated energy differences are undone as the phase of each ion returns to its starting position. At this point, the time spread of the bunch is compressed to its initial width, resulting in a maximum mass resolving power for that particular storage time. This presents an opportunity to fully mitigate the sinusoidal noise and return to the ideal mass resolving power value, regardless of the applied noise amplitude. It should be noted that for increased noise amplitude, the range of optimal storage times in a single sinusoidal period to extract the ions while achieving a high mass resolving power decreases. 

\begin{figure}[t]
\centering
\includegraphics[width=1.0\linewidth]{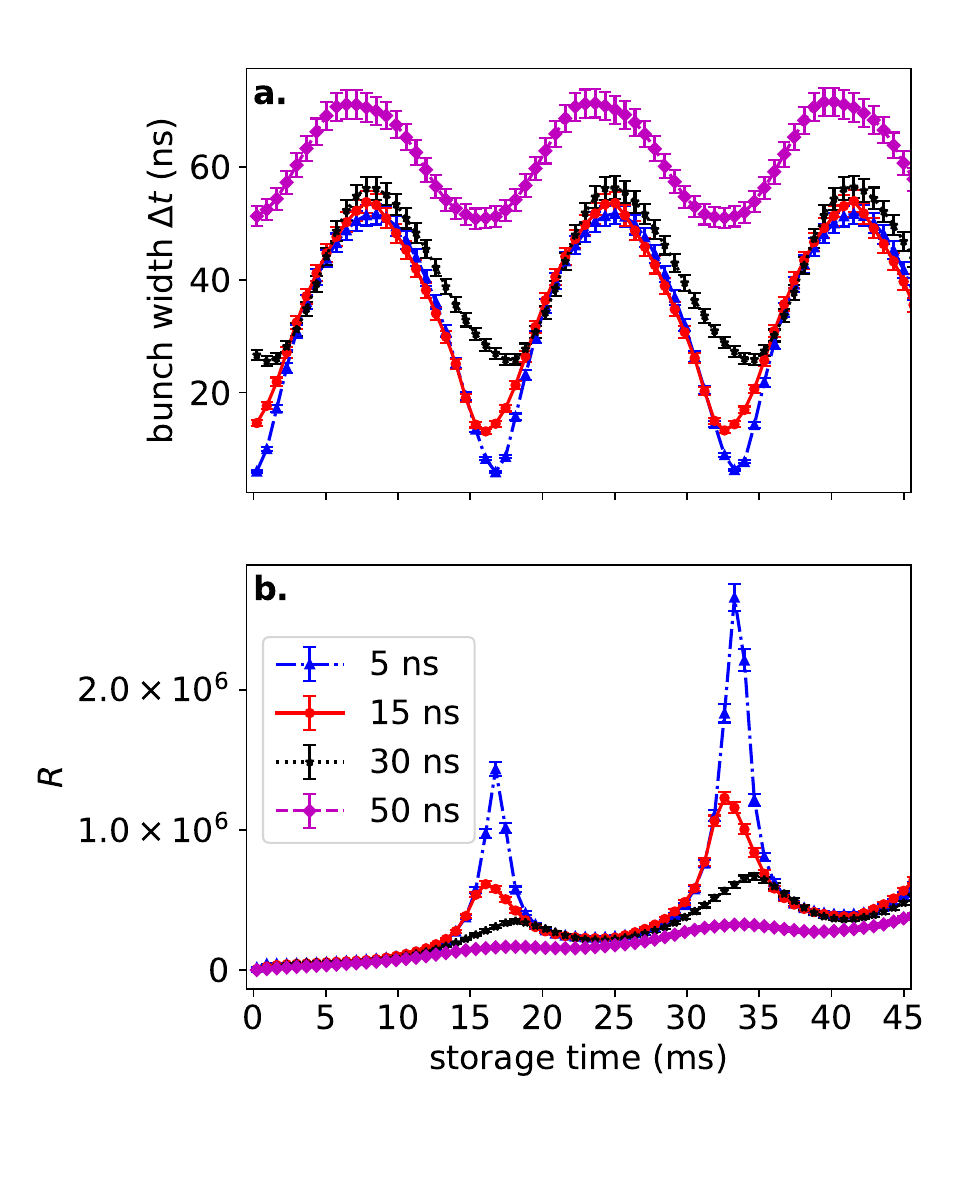}
\caption{(a.) The change in the bunch width and (b.) the mass resolving power $R$ of stored ions as a function of storage time for various initial bunch widths assuming a 60~Hz sinusoidal noise amplitude of 0.5~V and a random initial phase per ion. The blue curve is the same as depicted in Fig.~\ref{fig:Sinus1}.}
\label{fig:Sinus2}
\end{figure}

In order to mass separate ions of interest with short half-lives, the MR-ToF device is required to reach high mass resolving powers in relatively short storage times. Per Equation~\ref{Eq: ResolvingPower}, the storage time needed for a given mass resolving power is proportional to the initial temporal bunch width of the stored ions $\Delta t_0$. Many existing MR-ToF devices store ion bunches with initial bunch widths on the order of tens of nanoseconds, far exceeding the narrow time spreads simulated to this point. It is therefore interesting to study the influence of sinusoidal noise on ion distributions with larger initial bunch widths. The results of these simulations are shown in Fig.~\ref{fig:Sinus2} for a noise amplitude of 0.5~V. For larger bunch widths, the relative change in the revolution period of stored ions, for a set sinusoidal noise amplitude, is less pronounced. Despite larger initial bunch widths being impacted less by the sinusoidal noise, smaller initial bunch widths still reach a given mass resolving power in shorter storage times. Observing the oscillations in the time spreads shown in Figures~\ref{fig:Sinus1} and~\ref{fig:Sinus2} requires that the ion bunches are not phase locked to the 60~Hz noise signal, the storing of ion with relatively narrow time spreads ($\lesssim50$~ns), and some grounded potential region existing in the device\footnote{The effects shown in Figures~\ref{fig:Sinus1} and~\ref{fig:Sinus2} can be removed entirely by imposing the same sinusoidal noise on all grounded electrodes within the device in addition to the mirror electrodes.}.

\begin{figure}[t]
\centering
\includegraphics[width=1.0\linewidth]{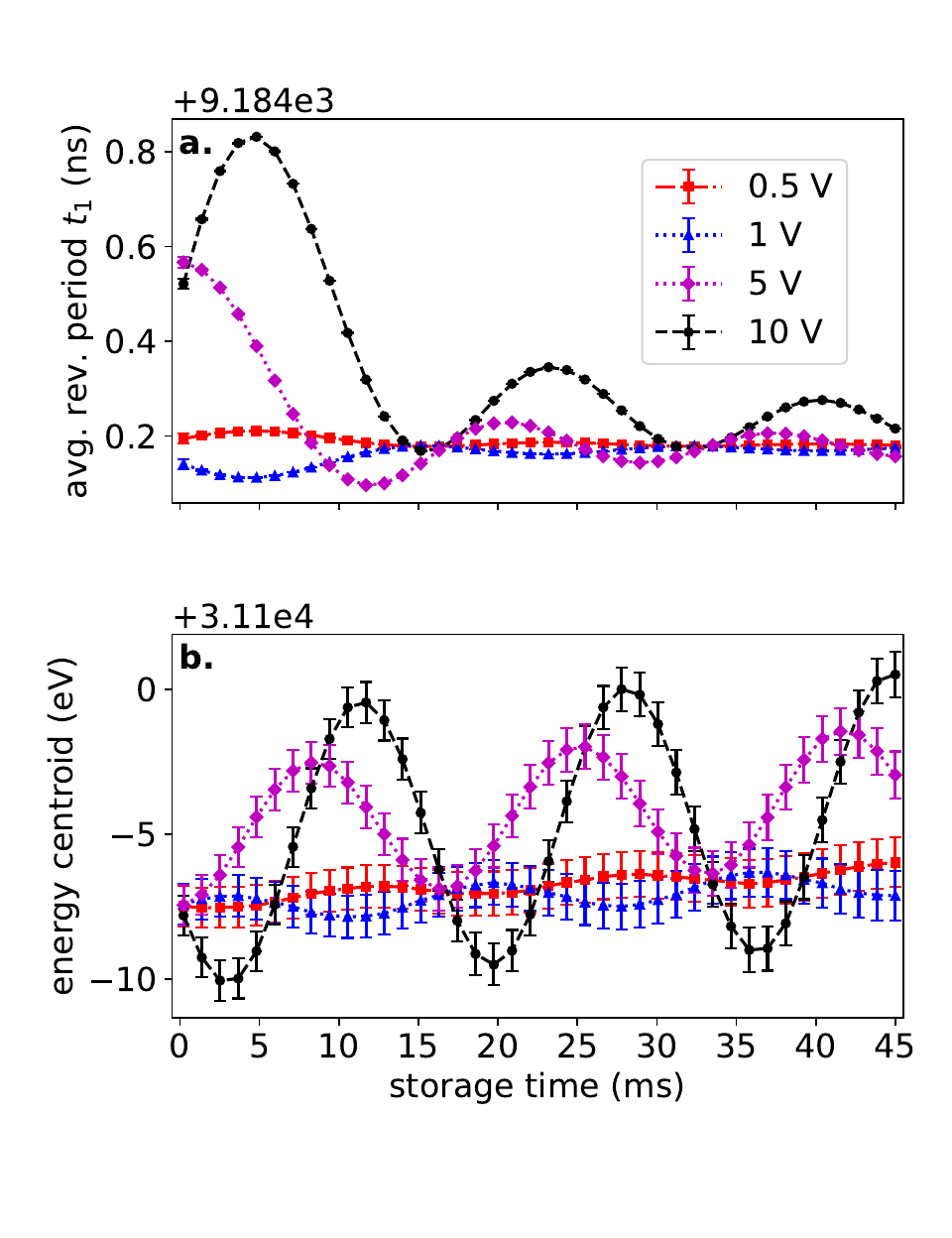}
\caption{(a.) The periodic change in the average revolution period, $t_1$, (taken as the total storage time divided by the number of revolutions) and (b.) the energy centroid as a function of storage time for 5~ns bunch widths under various 60~Hz sinusoidal noise amplitudes, with an identical phase per ion. In the case of panel a, the calculated error bars are smaller than the size of the data points.}
\label{fig:Sinus3}
\end{figure}

\textbf{Sinusoidal noise (phase-locking)--}
In the case where the extraction from the Paul trap can be phase-locked to the 60~Hz sinusoid, all stored ions will experience (nearly) the same initial phase of the noise. Simulations are performed utilizing this phase-locked approach. As before, all mirror electrodes are given the same initial phase of the sinusoid (assuming the same ground connection)\footnote{Simulations are also done for both the phase-locked and non-phase-locked case with 0.5~V noise amplitude and 5~ns bunches for different phase offsets in each electrode pair, thus assuming different ground connections for the mirror electrode power supplies. Similar results to those in Figs.~\ref{fig:Sinus1}-\ref{fig:Sinus3} are found, with the exception that for phase-unlocked tests, varying oscillation amplitudes in the bunch widths are observed.}. However, now, this initial phase is kept fixed for each individually stored ion. We again store ions for 5000 revolutions and simulate each ion's flight time individually. Each ion, starting at the same phase, experiences the same difference in potential between the mirrors and the grounded regions. The mass resolving power curve is hence identical to the one without any sinusoidal noise added across all simulated sinusoidal noise amplitudes up to 10~V, which is significantly exceeding the noise amplitude of the foreseen power supplies of $<0.5$~V. This stability against phase-locked sinusoidal noise is attributed to the energy tolerance of the FRIB design ($>$50~eV based on the desired $m/\Delta m$), which far exceeds the amplitude of oscillations in the energy centroid of stored ion bunches. The oscillations in the average revolution period (taken as the total time of flight divided by the number of revolutions), as well as the energy centroid for a given storage time, is shown in Figure~\ref{fig:Sinus3}. The apparent average increase in the energy centroid over time is thought to be a consequence of computational errors, considering it can be severely reduced by improving the grid resolution in SimIon. For each simulated curve displayed in Figure~\ref{fig:Sinus3}, a different locked initial phase is utilized, which all yield a mass resolving power in line with the ideal stable curve. Similar tests of different locked phases for various initial bunch widths (up to 100~ns) also yield similar mass resolving powers to their respective stable cases. The lack of an optimal phase is expected, considering the observed behaviors in Figs.~\ref{fig:Sinus1}-~\ref{fig:Sinus3} depend solely on whether ions see the same or different mirror potentials relative to a grounded region of the device.

\textbf{Gaussian white noise--}
To simulate the impact of voltage instabilities occurring at time scales far shorter than the revolution period, we add Gaussian white noise to the ideal mirror electrode potentials. In the simulations, at every single time step, a new potential is applied to the individual mirror electrodes, which we calculated using the polar form of the Box-Mueller transformation~\cite{10.1214/aoms/1177706645} with a standard deviation $\sigma$, following the same procedure as in Ref~\cite{MAIER2023167927}. We used an ion bunch with an initial bunch width of 5~ns composed of two masses (150 ions each) with a mass ratio of $m/\Delta m=10^6$. Ions are simulated one at a time, and their time of flight is recorded at a storage time of 12~ms (sufficient for $R=10^6$ with 5~ns bunch widths). Even under a noise with $\sigma=5$~V, far exceeding that of the foreseen power supplies ($\sigma<0.15$~V), the ability to mass separate is fully maintained. This robustness arises because the fast variation in the electrode potentials from Gaussian white noise causes simulated ions to experience, on average, the same mirror potentials. 

\subsection{\label{subsec:VoltageMit}Mitigation of impacts}
As shown in Sections~\ref{subsec:LT} and~\ref{subsec:ST}, the ion properties $t$ and $\Delta t$ can be shifted and broadened, respectively, by the presence of voltage instabilities. Extensive work has been done to enhance the stability of  MR-ToF systems~\cite{doi:10.1063/1.5104292, WIENHOLTZ2020348, Fischer2021HVstab} and modern power supplies~\cite{PASSON2025101818, 10.1063/5.0218649}. We will provide a brief overview of these efforts and discuss their relevance to the FRIB MR-ToF MS.

\textbf{Ultra-stable high-voltage power supplies--}
A targeted effort to improve the stability of high-voltage power supplies for use in MR-ToF devices was tested at low voltages on the RIKEN MR-ToF device~\cite{SCHURY201439} utilizing negative active feedback loops with emulated operational amplifiers; see detailed discussion in Ref.~\cite{doi:10.1063/1.5104292}. Such regulation blocks led to fast and reproducible settling times (1~ppm precision after 1 minute), remained stable at the level of 1~ppm for 1~hour and, in principle, can be made to to work up to and exceeding 15~kV~\cite{doi:10.1063/1.5104292}. 

\textbf{Passive and active stabilization--} 
Ensuring that subsequent ion bunches experience similar potentials within the MR-ToF device is crucial for achieving a high mass resolving power. This can be aided by passive and active stabilization, as detailed in Refs.~\cite{doi:10.1063/1.5104292, WIENHOLTZ2020348, Fischer2021HVstab}. Low-pass frequency filters allow for the suppression of sinusoidal and Gaussian white noise~\cite{WIENHOLTZ2020348, Fischer2021HVstab}. Considering that most MR-ToF devices operate with stored beam energies $<3$~keV, implementing low-pass filters does not pose safety concerns regarding the required capacitors. This is not the case for high-voltage MR-ToF MS. However, as demonstrated in section~\ref{subsec:ST}, short-term voltage fluctuations are not expected to be of concern for mass separation. 

On the other hand, long-term voltage fluctuations discussed in Section~\ref{subsec:LT} will considerably worsen the mass resolving power of the MR-ToF device and should be actively stabilized. This stabilization is possible via proportional-integral (PI) controllers in a fast feedback loop. The applied potentials need to be regularly read out as accurately as possible with stable voltage dividers and high-precision multimeters. Active adjustments can be made to correct a given electrode potential (almost) back to its set value to help mitigate long-term voltage drifts. A long-term stability better than 1 ppm has been demonstrated for the $<5$~kV power supplies employed for state-of-the-art low-voltage MR-ToF devices, see e.g.~\cite{WIENHOLTZ2020348, Fischer2021HVstab}. As demonstrated in Refs~\cite{10.1063/5.0218649, PASSON2025101818} for the purpose of collinear laser spectroscopy, a similar performance is also achievable for the high-voltage power supplies required for FRIB's 30~keV MR-ToF device. 

Thanks to the rapid progress in high-voltage power supply stabilization in recent years, high-voltage MR-ToF MS are expected to achieve a mass resolving power on par with the current state-of-the-art low-voltage setups that reach and slightly exceed $R>10^6$~\cite{ROSENBUSCH20231e6, Mardor20211e6}, while substantially improving the ion flux. With further anticipated improvements to high-voltage power supplies and newly developed mitigation strategies, the mass resolving power of FRIB's MR-ToF MS might one day approach the ideal-case value of $10^7$.  

\section{\label{sec:Temperature}Thermal expansion of the spectrometer}
\subsection{\label{subsec: Tempintro}Description and simulated impacts}
Changes in the ambient temperature will cause mechanical expansion or contraction of the MR-ToF device, resulting in shifts in the time of flight and a broadening of the measured bunch width $\Delta t$ when summing up many measurement cycles. Without any temperature stabilization, the ambient temperature at the foreseen location of the MR-ToF MS in the experimental hall varies as much as $\pm3$~K based on temperature measurements spanning an entire year. For time scales of a few days, the temperature varies by $\pm 1$~K. Assuming that the MR-ToF electrodes are constructed from Invar (thermal expansion coefficient~$\approx1\cdot10^{-6}/$~K~\cite{warlimont2018springer}) and isolated from the vacuum chamber, a 1 K temperature change would contribute a $\pm$1~ppm change in the MR-ToF dimensions. For titanium or stainless steel (thermal expansion coefficient~$\approx1\cdot10^{-5}/$~K), this would be $\pm$10~ppm.  

\begin{figure}[t]
\centering
\includegraphics[width=1.0\linewidth]{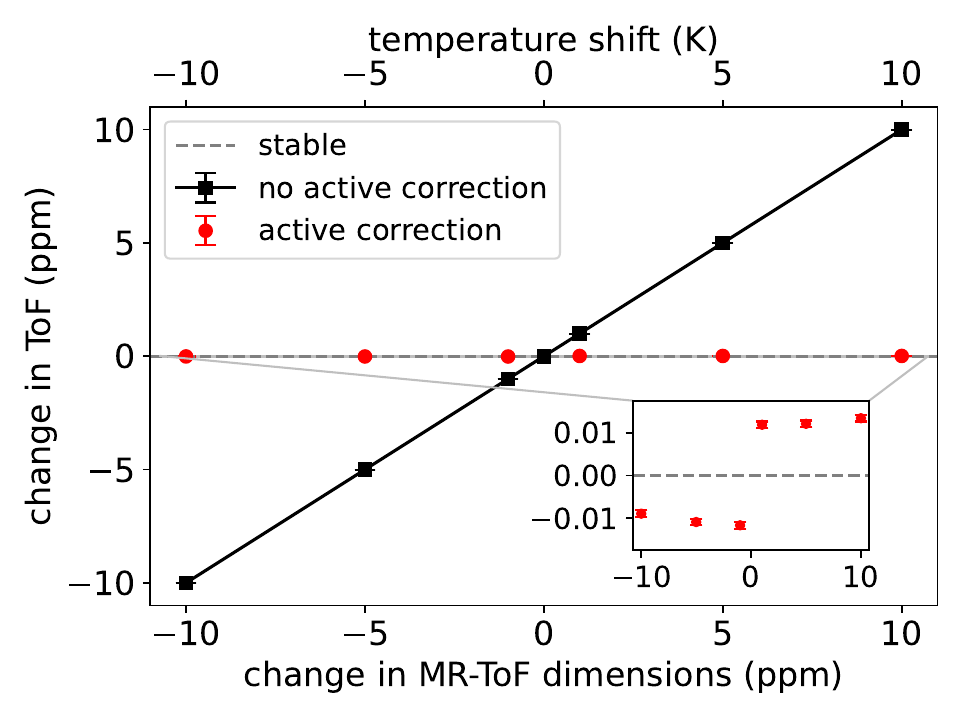}
\caption{The simulated change in the time of flight (ToF) centroid of stored ions after 25 revolutions as a function of change in MR-ToF dimensions (in units of parts-per-million) when no active correction voltage is applied (solid black line) and when a +2.571V/(ppm~ToF) correction is applied to the pickup electrode (P) shown in Fig.~\ref{fig:Schematic} (red data points) compared to perfect stability (dashed gray line). The corresponding temperature shift (in~K), assuming the MR-ToF setup is constructed out of Invar, is indicated on the upper horizontal axis. The error bars are smaller than the individual data markers.}
\label{fig:Temp}
\end{figure}

The impact of temperature fluctuations on the mass resolving power can be simulated by varying the dimensions of the MR-ToF device by some parts-per-million via the positioning scaling factor in SimIon. For this test, the MR-ToF device depicted in Fig.~\ref{fig:Schematic} is assumed to be mounted on the same structure, meaning that the entire device will expand or contract together. Ion bunches undergo 25 revolutions before having their time-of-flight centroid recorded for a given parts-per-million expansion/contraction. These simulations show a 1~ppm change in time of flight per 1~ppm change in the MR-ToF dimensions, see the black curve in Figure~\ref{fig:Temp}. Assuming the device is constructed out of Invar, the temperature change of $\pm 1$~K results in a $\pm1$~ppm change in time of flight. For a setup made from titanium or stainless steel, the same temperature change would result in a $\pm10$~ppm change in time of flight. This is in line with $<3$ keV MR-ToF MSs that experimentally observe a 5 to 10 ppm time of flight change per 1~K~\cite{SCHURY2013537, SCHURY201439, MehlhornBSc, PhysRevC.99.064313.trccorr}. Temperature instabilities of $\pm3$~K or $\pm0.5$~K reduce the long-term mass resolving power of FRIB's 30 keV MR-ToF device to $R\approx 7.5 \cdot 10^4$ or $R\approx4.5\cdot 10^5$, respectively. As a result, temperature fluctuations during the measurement time severely impact the mass resolving power and must be addressed.

It is also important to assess whether a major temperature change in the experimental hall housing the MR-ToF MS could affect the required tuning of the mirror electrode potentials to maintain high mass resolving power for a largely different, but fairly stable, temperature. No increased broadening in $\Delta t$ for one single stored ion bunch is observed for changes in the MR-ToF dimensions up to as high as $\pm1000$~ppm, assuming the length changes are uniform along the entire device. Hence, even after a severe temperature change, a retuning of the device will not be required. 

\subsection{\label{sec:TemperatureMit}Mitigation of impacts}
\textbf{Environmental stabilization--}
Considering the worsening of mass resolving power due to ambient temperature fluctuations, the MR-ToF device should be housed in a dedicated temperature-stable environment to reduce the expansion/contraction of the structure and be well isolated from the vacuum chamber. Additionally, temperature sensors could be installed for an active temperature stabilization of the environment surrounding the setup~\cite{MehlhornBSc}. Furthermore, the high-voltage power supplies and the voltage dividers should also be housed within a temperature-stabilized environment, since temperature fluctuations also affect the applied potentials. Like this, the stability in the ambient temperature surrounding the MR-ToF device could be significantly improved down to a few tens of mK, as recently demonstrated~\cite{MehlhornBSc}. For a mass resolving power above $3\cdot 10^6$ and assuming construction of the setup out of Invar, a stability of $\pm50$~mK is required. For $R\approx 10^7$, we would require a stability of $\pm2$~mK.  

\textbf{Correction via regular calibration measurements--}
For mass spectrometry, the effects of long-term voltage and temperature drifts are significantly reduced when performing time-of-flight measurements of reference ions that are regularly interleaved with those of the ions of interest. During the analysis process, these reference measurements are used to correct for the drift in the time centroid, achieving a significantly higher mass resolving power~\cite{PhysRevC.88.011306.tofcorr, DICKEL2015172, PhysRevLett.120.062503, PhysRevC.97.064306, WOLF2013123, Fischer2018, PhysRevC.99.064313.trccorr}. It has been shown that this is possible for both isobaric and non-isobaric reference ion species.  Very often, a contaminant species is present in the rare-isotope beam together with the ions of interest. If its rate is sufficiently high, the contaminant species can be used for calibration, reducing the number of reference measurements required. If the mass of the contaminant species is well-known and close to the one of the ions of interest, the need for interleaved calibration measurements is fully removed. 

For mass separation, a similar procedure to mitigate remaining long-term drifts has not been reported to date since active stabilization with a feedback loop requires destructive measurements that would impede the delivery of the separated beam. In this case, regular time-of-flight measurements would be required with corresponding adjustments for the switch timings of the high-voltage switches used for contaminant removal during the measurement, which is less ideal. As an example, when the purified beam is provided to downstream experiments, the ion detector needs to be retracted, but to perform calibration measurements, it needs to be on-axis. Repeated switching of the ion detector on and off axis will ultimately lead to mechanical wear and eventual failure of the linear drive mechanism. In mass separation mode, there exist a few options to monitor the time-of-flight signal and use it as input for an automatized correction of the switch timings relevant for contaminant removal. Firstly, a second, fixed, ion detector could be located slightly off-axis and a kicker electrode located between the two ion detectors could be used to guide the beam onto this second detector whenever calibration measurements are performed. This avoids frequent switching of the first ion detector between its on- and off-axis location. Secondly, in cases where sufficiently many ions are stored in the device, the image charge signal on the dedicated pickup electrode within the MR-ToF device can be used to monitor the ion time of flight~\cite{doi:10.1021/acs.analchem.7b02797}. In the case where the rate of the delivered beam is too low to produce a sufficient pickup signal, it could be investigated whether a high-rate non-isobaric stable contaminant could be additionally injected into the MR-ToF device. This high-rate ion species would produce a strong pickup signal that could be utilized for long-term corrections. Since its mass is far from the mass of the ions of interest, space-charge effects will be minimal~\cite{MAIER2023168545}, and this far-lying calibration ion species can be easily removed prior to the extraction of the purified ion beam. Once the time of flight is recorded either with regular calibration measurements on a time-of-flight detector or via the pickup of image charge signals, it can be used to apply corrections to the switch timings of the HV switches employed for contaminant removal. 

Alternatively, and likely the better solution, the potentials applied to electrodes in the MR-ToF device could be slightly changed to actively counteract the observed shift in the time of flight. To demonstrate this newly suggested procedure in simulation, we utilize the pickup electrode (P) shown in Fig.~\ref{fig:Schematic} to correct for the shifts in the time centroid from the temperature fluctuations discussed in Section~\ref{subsec: Tempintro} and shown in Fig.~\ref{fig:Temp}. The pickup electrode is found to yield a 1~ppm change in time of flight per 2.571~V applied. Given that the required voltage corrections are less than a few volts for the pickup electrode, high-precision, low-voltage power supplies which have better than $1$~mV precision and stability in their set value can be used (e.g. Stahl-Electronics UM/BS/BSA series). The red data points highlighted in the inset of Figure~\ref{fig:Temp} show the improvement in the time centroid stability when implementing the relevant potential correction for a given temperature change. We find an ability to stabilize the time centroid to a precision of $\pm0.01$~ppm when setting the correction voltage to 1~mV precision. This stability in the time centroid of stored ion bunches would enable a mass resolving power $R\approx 7.5 \cdot 10^6$, which almost approaches the mass resolving power in ideal conditions. Accuracy in the set voltage that is better than 1~mV  will enable even higher mass resolving powers. 

We additionally simulated an order of magnitude higher change in the MR-ToF dimensions, assuming the device would be constructed from titanium, and found similar success. Assuming an ability to monitor the time of flight, active time centroid adjustments would mitigate the worsening from all forms of long-term drifts to an extent that resolving powers well exceeding $10^6$ could be realized. Instead of monitoring the time of flight, one could also measure the ambient temperature to determine the voltage corrections needed to counteract the temperature drifts. This would require a characterization of the time-of-flight changes based on changes in ambient temperature with stable ions produced in an offline ion source prior to the actual mass separation.

\section{\label{sec:Further}Further expected non-ideal conditions}
\subsection{Voltage inaccuracies}
Once a good potential distribution is found for an MR-ToF device, maintaining it after a restart of the setup requires not only high precision, but also highly accurate power supplies. The set values of commercially available 60~kV power supplies (e.g FUG HCP series) can be off by as much as $\pm1000$~ppm from the desired input when reading the voltage off of the front panel of the supply. When reading with a fine potentiometer, this can improve to roughly $\pm10$~ppm. For the active stabilization, we will use high-precision voltage dividers~\cite{PASSON2025101818, 10.1063/5.0218649} and read out the voltage via a DMM7510 multimeter that has a resolution of $\pm1$~ppm and an accuracy of approximately $\pm22$~ppm over the course of two years. If the voltage dividers are not well calibrated (e.g their calibration factor changes over time for instance due to a change in temperature or an aging of the resistors), the accuracy of the potential applied to the electrodes might further suffer. Hence, it is desirable to assess the impact of inaccuracies in the electrode set values on the mass resolving power $R_{\mathrm{inf}}$. To simulate this, we deviate the set values of the mirror electrodes by up to $\pm1000$~ppm from their optimal value in Table~\ref{tab:MRToFSetting} in the appendix and record the revolution period $t_1$ and broadening per revolution $\Delta t_1$ of an ion bunch after 25 revolutions in the device to determine $R_{\mathrm{inf}}$. We find that mirror electrodes 1-5 can be individually offset by up to $\pm1000$~ppm (assuming the potentials of the other 8 electrodes are optimal) without worsening $R_{\mathrm{inf}}$ below $10^7$. Mirror electrodes 6-8 can be offset by up to $\pm150$~ppm and $\pm500$~ppm and still yield $R_{\mathrm{inf}}>10^7$ and $R_{\mathrm{inf}}>3 \cdot10^6$, respectively. The outermost electrode (\#9) can be offset by up to $\pm100$~ppm and $\pm250$~ppm and still yield $R_{\mathrm{inf}}>10^7$ and $R_{\mathrm{inf}}>3 \cdot10^6$, respectively. From these results, we foresee that the resolving power will not be impacted by the foreseen voltage inaccuracies. 

\subsection{Realistic high-voltage switches}
Practically, high-voltage switch circuits exhibit some non-zero time constants that will slightly alter the potentials seen by stored ions during the first few revolutions. For ion storage/extraction, the drift tube electrode in Fig.~\ref{fig:Schematic} is switched between $\approx$~18.9~kV and ground. Up to this point, simulations have assumed that the change between these potentials occurs instantaneously. To model a realistic high-voltage switch, we adopt the procedure discussed in Refs.~\cite{MAIER2023167927, MAIER2025170365} including the switch response function shown there. No worsening in the mass resolving power is observed in our simulations of over 125 ms of storage time.

\begin{figure}[t]
\centering
\includegraphics[width=0.75\linewidth]{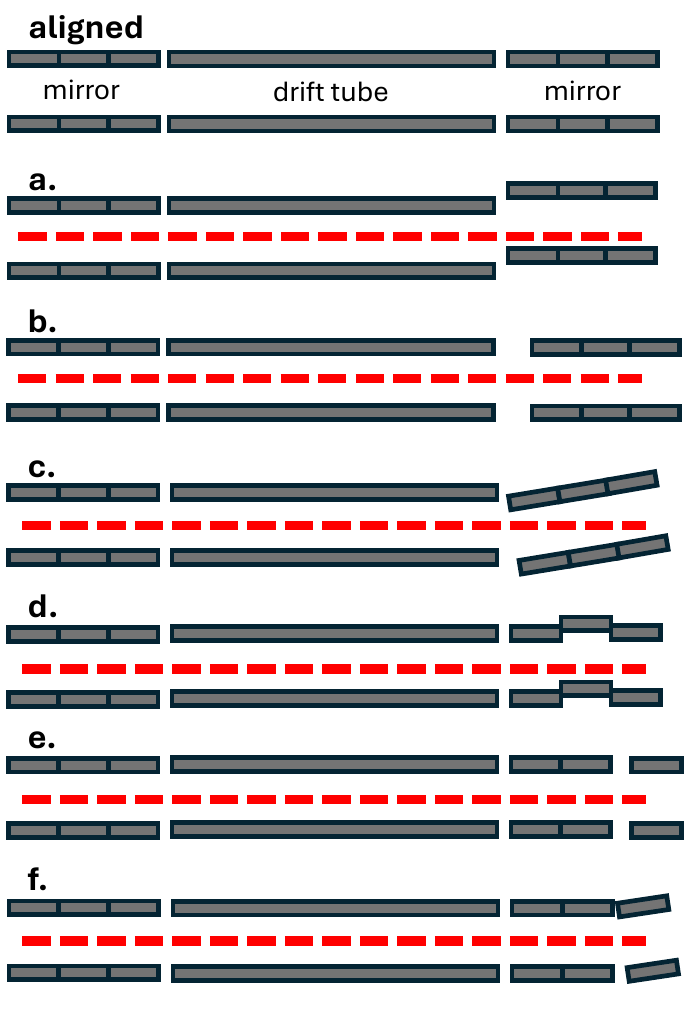}
\caption{A simplified schematic of an MR-ToF device with only three electrodes per mirror illustrating exaggerated forms of misalignments/asymmetries simulated for the case when one full mirror section is offset by either a shift (\textbf{a.} and \textbf{b.}) or tilt (\textbf{c.}), or for when an individual electrode is offset by either a shift (\textbf{d.} and \textbf{e.}) or tilt (\textbf{f.}) relative to the remaining fixed setup and the ion optical axis (red dashed line).}
\label{fig:Angle}
\end{figure}

\subsection{Electrode misalignments and mirror asymmetries}
Misalignments between the mirror electrodes and/or entire mirror sections will alter the trajectories of stored ions and worsen the resolving power. These misalignments are determined by the machining tolerance of the electrode/insulator configuration and the alignment precision. State-of-the-art laser alignment and CNC machining are expected to align individual components of the FRIB design to below tens of micrometers. Simulating such small misalignments in SIMION requires a very fine grid resolution (defined as the number of grid units (gu) per millimeter). Up until this point, the MR-ToF MS was simulated at a resolution of 25~gu/mm to minimize computational errors. We have therefore exploited the cylindrical and axial symmetry of the device to reduce the cost of the simulations. To simulate misalignments, these symmetries must be broken. As a result, we must sacrifice a factor of 5 in grid resolution and can as a result only simulate misalignments $>0.2$~mm. For a resolution of 5~gu/mm, and in a perfectly aligned setup, the mass resolving power $R_{\mathrm{inf}}$ of the FRIB design decreases from $2\cdot10^7$ to $6.5\cdot10^6$, a factor 3 worsening purely from computational errors.

While we cannot practically simulate all possible misalignments within the MR-ToF device, testing the various cases illustrated in Figure~\ref{fig:Angle} will give insight into the impacts of various general types of misalignment on the mass resolving power. For the panels \textbf{a.}, \textbf{b.}, and \textbf{c.} of Fig.~\ref{fig:Angle}, an entire mirror, consisting of the grounded shielding electrode, pickup/deflector electrode and nine mirror electrodes, is either vertically or horizontally translated or tilted with respect to the ion optical axis (red dashed line) defined by the Paul trap and injection optics. We simulate 300~ions with an initial bunch width of 5~ns performing 5000 revolutions in the device, corresponding to a storage time of roughly 50~ms. Based on this dataset, we determine $R_\mathrm{inf}$. Even for worst-case shifts of 0.6 mm or tilts of 0.1 degree, corresponding to $\approx 0.6$~mm shifts in position of the outer edge of the mirror relative to the ion optical axis, $R_{\mathrm{inf}}\gtrsim 6\cdot10^5$. We also shift an individual mirror electrode, both horizontally and vertically, or tilt it as shown in panels \textbf{d.}, \textbf{e.}, and \textbf{f.} of Fig.~\ref{fig:Angle}. For shifts of 0.2 mm or tilts of 0.2 degrees, which correspond to $\approx 0.2$~mm shifts in position relative to the ion optical axis, $R_{\mathrm{inf}} \gtrsim 6\cdot10^5$ irrespective of the chosen mirror electrode. Larger misalignments are tested for the electrode nearest the turnaround point of the ions (electrode \#7). Even for tilts of $1.3$~degrees for electrode \#7, corresponding to $\approx1$~mm shifts in position relative to neighboring electrodes, $R_{\mathrm{inf}}\gtrsim2.5\cdot10^5$. Across all tests, no ion losses are experienced. 

Misalignments of the full MR-ToF device with respect to the ion optical axis are also tested. For these simulations, a grid resolution of 10~gu/mm is possible. For a perfectly aligned setup, the mass resolving power for this grid resolution is $R\gtrsim1.5 \cdot 10^7$. For tilts of the entire device about the MR-ToF middle plane by $\approx 0.1$~degrees, resulting in a shift of both mirrors by 1~mm with respect to the ion optical axis, the mass resolving power is reduced to $R\approx3.5 \cdot10^6$. For vertical shifts of the entire device by 1~mm with respect to the ion optical axis, the mass resolving power is $R\approx2.5 \cdot10^6$. Horizontal shifts of similar magnitude only slightly alter the flight time of the ions at the moment they are stored in the MR-ToF device, which is not having an impact on the mass resolving power. In practice, the misalignment of the entire MR-ToF device with respect to the ion optical axis will be far lower than in our simulations and can additionally be compensated for by steering of the ions upon injection into the MR-ToF device.

The relatively high mass resolving power of the design across various types of misalignments is partly due to the small transversal beam envelope of the stored ions prepared by the upstream Paul trap relative to the large inner diameter of the MR-ToF device. As a result, even for a large misalignment, stored ions do not experience drastically different potentials compared to the perfectly aligned case, and the mass resolving power remains high. Considering the misalignment in the experimental setup is expected to be far smaller than the cases tested in simulation, high mass resolving powers well above $10^6$ are anticipated.

Moreover, the potentials applied to the mirror electrodes can be retuned to improve the mass resolving power in the presence of said misalignment. To simulate this, we store the same 300~ions utilized in the initial misalignment tests for 1000~revolutions in the setup where mirror electrode \#7 is tilted by $1.3$~degrees, which yields the most significant worsening to the mass resolving power across all simulated misalignments above ($R\gtrsim 2.5\cdot10^5$). When reoptimizing the potentials applied to the mirror electrodes following the same simulation routine as discussed in Ref.~\cite{MAIER2023168545}, we could improve the mass resolving power by an order of magnitude to $R\gtrsim2.5\cdot10^6$. This is approaching the resolving power found in the perfectly aligned case for the same grid resolution of 5~gu/mm. This result shows that mitigating the worsening of the mass resolving power due to misalignments may be possible by a retune of the MR-ToF mirror potentials.

\begin{figure}[t]
\centering
\includegraphics[width=1.0\linewidth]{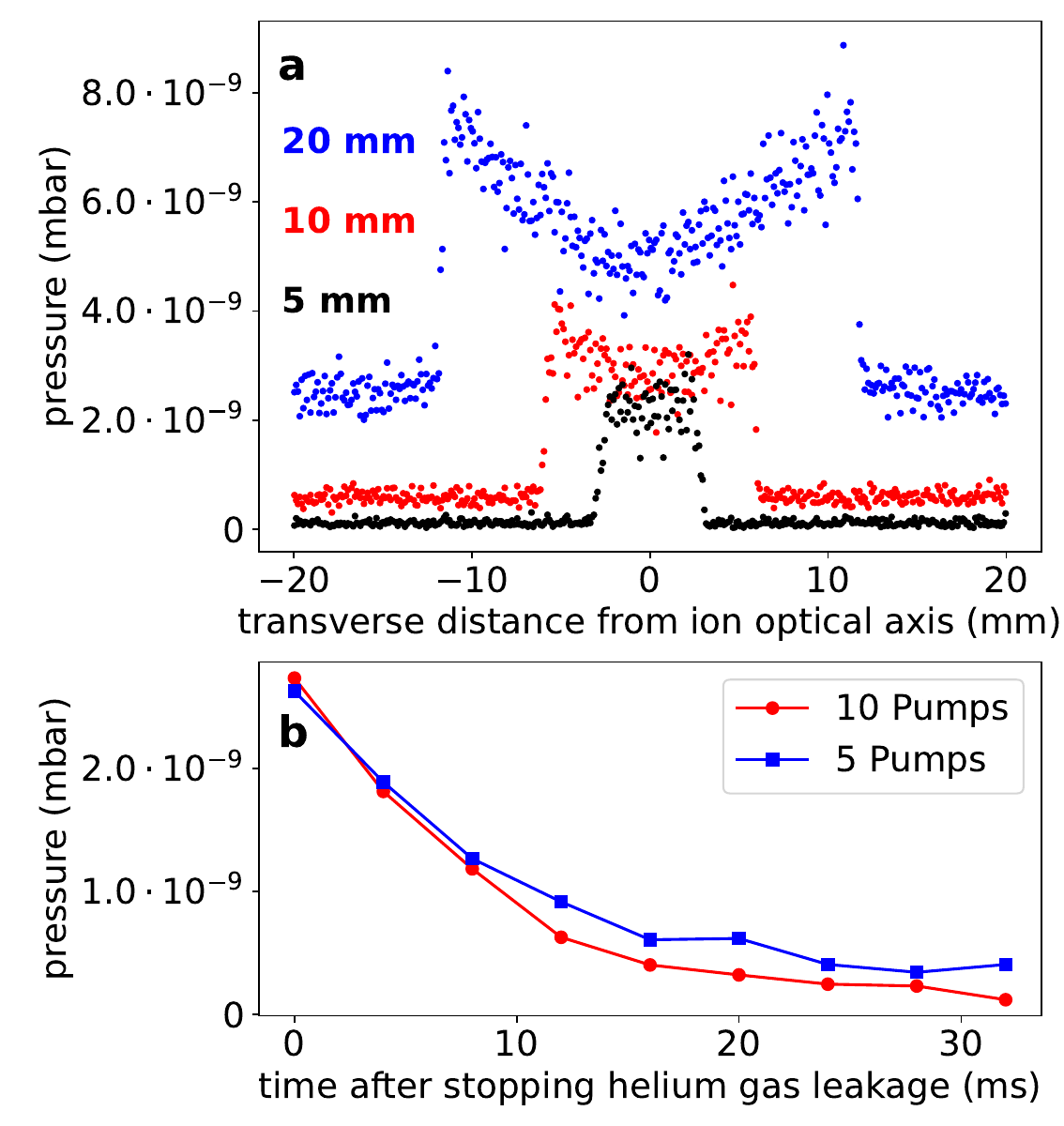}
\caption{(a.) Simulated helium pressure as a function of transverse distance from the ion optical axis at the entrance to the MR-ToF device for various iris apertures (labeled vertically as they appear). (b.) Simulated on-axis helium pressure at the entrance of the MR-ToF device as a function of time after stopping gas leakage into the setup.}
\label{fig:Helium}
\end{figure}

\subsection{Helium leakage into the MR-ToF device}
 A helium buffer-gas-filled Paul trap cooler-buncher is required upstream of the MR-ToF device to optimally prepare ion bunches. Helium atoms can escape from the Paul trap and pass through the injection optics toward the MR-ToF device, potentially causing a decrease in the mass resolving power due to collisions between the helium atoms and stored ions. The importance of ultra-high vacuum in an MR-ToF device has been reported by multiple state-of-the-art setups~\cite{DICKEL2015172, REITER2021165823, ROSENBUSCH20231e6}. We have thus far assumed a perfect vacuum for all our simulations. While the lack of knowledge of the velocity dependent collision cross sections prohibits reliable determinations of the worsening in mass resolving power due to non-perfect vacuum, understanding the expected helium leakage into our setup and comparing this with existing setups will provide insight into what performance can be reached.

To assess the impact of the helium flow into the FRIB MR-ToF device, we performed Monte Carlo pressure simulations in Molfow~\cite{Kersevan:IPAC2019-TUPMP037} within the proposed Paul trap and injection optics discussed in Ref.~\cite{FRIBMRToF}. We assume a background pressure of $1\cdot10^{-7}$ mbar in the beamline sections before the Paul trap and after the MR-ToF device. It is also assumed that 70\% of the helium atoms reaching the boundaries at the interfaces — between the FRIB vacuum chamber and the Paul trap, and between the MR-ToF device and the downstream ion detector — are effectively pumped away\footnote{No major changes are seen when assuming that 50\% or 90\% of the helium atoms reaching the boundaries are pumped away instead of the 70\%.}. In the simulations, we ignore outgassing from the vacuum chamber walls or ion optical elements. The helium pressure in the high-pressure region of the Paul trap is  $\approx1 \cdot 10^{-2}$~mbar as required for optimal stopping of the injected ion bunch. 

Based on simulation results in Molflow, we designed injection optics to restrict the helium flow to the MR-ToF device; see Ref.~\cite{FRIBMRToF}. Two adjustable iris shutters, based on the design detailed in Ref.~\cite{Klink_2024}, will enable differential pumping between the Paul trap and the MR-ToF device.  When the irises are fully open (30~mm diameter), the simulated on-axis pressure at the entrance to the MR-ToF device, 843 mm away from the Paul trap, is $\approx 1 \cdot 10^{-8}$~mbar. To lower the pressure further, the iris apertures can be reduced down to 3~mm without any simulated ion losses. Panel \textbf{a} of Figure~\ref{fig:Helium} shows the simulated pressures as a function of the transverse distance from the ion optical axis at the entrance to the MR-ToF device for three iris aperture sizes: 5~mm (black), 10~mm (red) and 20~mm (blue). The pressure on the ion optical axis is significantly higher than the pressure several (tens of) mm off-axis. A sizeable flow of helium atoms still enters the MR-ToF device. The pressure increase slightly off-axis compared to on-axis is attributed to gas molecules colliding with the ion optics. 

While the irises with a 5 mm diameter enable us to achieve an on-axis pressure of $\approx3 \cdot 10^{-9}$~mbar at the entrance of the MR-ToF device, which is already below the pressure reported in existing MR-ToF setups, a further reduction in pressure can be achieved by pulsing the helium injection in the Paul trap~\cite{PhysRevLett.102.233004}. To investigate this, simulations are conducted monitoring the on-axis pressure at the MR-ToF entrance as a function of time after the helium gas leakage into the Paul trap was stopped. For this test, a 10~mm diameter for the adjustable iris shutters is chosen. The results of the simulations are displayed in panel \textbf{b} of Figure~\ref{fig:Helium}.  Approximately 30~ms after stopping the helium gas leakage, the pressure drops by an order of magnitude. Therefore, if the ions have sufficiently long half-lives, helium injection can be halted while the ions remain stored in the Paul trap for several tens of milliseconds. This approach enables reduced helium pressure in the MR-ToF device by the time the ions are transferred into it.
No significant improvement in pressure is seen in the simulations when increasing the number of 600 l/s turbo pumps installed between the Paul trap and the end of the MR-ToF device from 5 to 10. If sizeable outgassing from the chamber walls and ion optical elements is present, an increased number of turbo pumps could, however, be useful in the experiment. 

The achieved simulated pressures, even without pulsed helium injection, are lower than the experimentally reported values of MR-ToF devices that reach and slightly exceed mass resolving powers of $10^6$~\cite{DICKEL2015172,ROSENBUSCH20231e6}. We therefore expect that the pressure in the MR-ToF device will not be a limiting factor in the mass resolving power.

\section{Conclusion and Outlook}
We presented a simulated analysis of non-ideal conditions that we expect to encounter during the operation of the FRIB high-voltage multi-reflection time-of-flight mass separator and spectrometer (MR-ToF MS). To fully meet the goals outlined in Ref.~\cite{FRIBMRToF}, mass resolving powers $R=m/\Delta m>10^6$ are desired. While the FRIB design exceeds $R=10^7$ under simulated ideal conditions, some deterioration is expected due to voltage instabilities and inaccuracies, temperature fluctuations, and misalignments. When assuming state-of-the-art stabilization routines for the high-voltage power supplies, so a voltage stability of $\pm0.5$~ppm, the mass resolving power worsens to $8\cdot 10^5$ due to long-term voltage drifts. Short-term voltage fluctuations are found to be either of no concern for mass separation or able to be mitigated by locking the Paul trap extraction to the phase of the expected 60~Hz sinusoidal noise. Temperature fluctuations can be largely mitigated by environmental stabilization. Misalignments, even for large shifts of 0.2~mm, only worsen the mass resolving power to $\gtrsim 6\cdot10^5$. When employing existing techniques for mitigation of non-ideal conditions, we hence expect that the mass resolving power of FRIB's high-voltage MR-ToF MS will be on par with existing low-energy MR-ToF MSs approaching $R\approx 1\cdot 10^6$, while improving the ion flux by over two orders of magnitude. We furthermore explored the possibility of an active time centroid adjustment for the correction of remaining shifts from long-term voltage and temperature fluctuations. In simulations, this method is found to stabilize the ion time of flight to $\pm0.01$~ppm when assuming that the set voltage of ultra-stable low-voltage power supplies used for implementing the correction are accurate to 1~mV. This would allow for resolving powers that approach $7.5 \cdot 10^6$. This technique can be readily implemented in FRIB's highly selective and high-flux MR-ToF MS once it is commissioned. These new mitigation efforts, along with the other findings reported in this manuscript, may also prove valuable for improving the mass resolving power of existing state-of-the-art MR-ToF MSs. 

\section*{\label{sec:Acknowledgments}Acknowledgments}
This work was supported by the U.S. Department of Energy, Office of Science, Office of Nuclear Physics under Grant No. DE-SC0023633 (MSU) and DE-AC02-05CH11231. C.M.I. acknowledges support from the ASET Traineeship under the DOE award no. DE-SC0018362. This work was supported by the U.S. Army DEVCOM ARL Army Research Office (ARO) Energy Sciences Competency, (Electrochemistry or Advanced Energy Materials) Program award \# W911NF2510038. The views and conclusions contained in this document are those of the authors and should not be interpreted as representing the official policies, either expressed or implied, of the U.S. Army or the U.S. Government. This work was supported in part through computational resources and services provided by the Institute for Cyber-Enabled Research at Michigan State University.

\section*{\label{Appendix1}Appendix 1}
The potentials applied to the individual electrodes of the FRIB MR-ToF design are stated in Tab.~\ref{tab:MRToFSetting}. 
\begin{table}[h]
\centering
 \caption{\label{tab:MRToFSetting} The potentials applied to the FRIB MR-ToF mirror electrodes, the positions of which are shown in Fig.~\ref{fig:Schematic}.}
 \vspace{1mm}
 %\begin{ruledtabular}
 \begin{tabular}{cc}
   \hline
   \hline 
  electrode & potential (V)\\
 \hline
  1 (innermost electrode) &  -12\,908.4 \\
  2 &  -14\,348.2 \\
  3 & -1\,147.0 \\
  4 & 12\,771.5 \\
  5 &  21\,564.2 \\
  6 & 25\,329.3\\
  7 &  32\,384.3 \\
  8 &  36\,144.2 \\
  9 (outermost electrode) & 43\,340.5\\
  \hline
  \hline
 \end{tabular}
% \end{ruledtabular}
 \end{table}

%\section*{Author Contributions}

%\section*{\label{sec:Appendix}Appendix}

%\bibliographystyle{elsarticle-num} 

\bibliography{biblio}% Produces the bibliography via BibTeX.

\end{document}